\documentclass[useAMS,usenatbib]{mn2e}

\def\Lo{$L_{\sun}$}

\newcommand{\uJy}{\ensuremath{\mu{\rm Jy}}}

\usepackage{graphicx}
\usepackage{epstopdf}

\def \etal {\rm ~{\it \etal},~}

\def\apj {{\it Ap.~J.}}
\def\apjl {{\it Ap.~J.\ (Letters)}}
\def\apjs {{\it Ap.~J.\ Suppl.}}
\def\aj {{\it A.~J.}}

\def\aap {{\it Astr.~Ap.}}

\def\araa {{\it Ann.\ Rev.\ Astr.\ Ap.}}

\def\mnras {{\it MNRAS}}

\def\kms {\ifmmode{{\rm ~km~s}^{-1}}\else{~km~s$^{-1}$}\fi}
\def\lsun {\ifmmode{{\rm ~L}_\odot}\else{~L$_\odot$}\fi}
\def\deg {^{\circ} }

\newcommand{\Mgas}{$M_{\rm H2}$}
\newcommand{\Msun}{~M$_{\odot}$}

\def\eps@scaling{1.0}%
\newcommand\epsscale[1]{\gdef\eps@scaling{#1}}%
\newcommand\plotone[1]{%
 \typeout{Plotone included the file #1}
 \centering
 \leavevmode
 \includegraphics[width={\eps@scaling\columnwidth}]{#1}%
}%

\title[VLBI of ULIRG F00183-7111]{The radio core of the Ultraluminous Infrared Galaxy F00183-7111: watching the birth of a quasar}

\author[Norris et al.]
{Ray P. Norris $^{1}$, Emil Lenc$^{1}$, Alan L. Roy$^{2}$, Henrik Spoon$^{3}$ \\
$^{1}$ CSIRO Astronomy \& Space Science, PO Box 76, Epping, NSW 1710, Australia\\
$^{2}$ Max Planck Institut fur Radioastronomie, Germany\\
$^{3}$ Cornell University, Astronomy Department, Ithaca, NY 14853, USA}

\begin{document}


\pagerange{\pageref{firstpage}--\pageref{lastpage}} \pubyear{2009}

\maketitle

\label{firstpage}

\begin{abstract}

F00183-7111 is one of the most extreme Ultra-Luminous Infrared Galaxies known. Here we present a VLBI image which shows that F00183-7111 is powered by a  combination of a radio-loud Active Galactic Nucleus surrounded by vigorous starburst activity. Although already radio-loud, the quasar jets are only 1.7 kpc long, boring through the dense gas and starburst activity that confine them. We appear to be witnessing this remarkable source in the brief transition period between merging starburst and radio-loud ``quasar-mode'' accretion.

\end{abstract}

\begin{keywords}

radio continuum: galaxies --- galaxies: evolution  --- galaxies: ULIRG --- galaxies: active --- VLBI

\end{keywords}

\section{Introduction}
\label{intro}
Ultraluminous infrared galaxies (ULIRGs) are a class of galaxy with a bolometric luminosity in excess of $10^{12}$ $L_{\odot}$ which were first uncovered by the IRAS satellite more than 20 years ago \citep{Aaronson:1984p29003, Houck:1985p29887, Allen:1985p29041}. A variety of observations suggest that they represent a transitional stage in which gas-rich spirals are merging to form a dusty quasar \citep{Armus:1987p29090,Sanders:1988p29218, Veilleux:2002p29465, Spoon:2009p29237}.   According to the scenario by \citet{Sanders:1988p29218}, the merger of two normal spirals fuels a pre-existing quiescent black hole and triggers a powerful nuclear starburst.This intense starburst activity causes the high infrared luminosity, and generates strong starburst-driven winds which will eventually blow away the enshrouding dust and lay bare the quasar core, depleting  the dust and gas to form an elliptical galaxy \citep{Dasyra:2006p29091, Dasyra:2006p29094}.   This scenario is supported by merger simulations \citep[e.g.][]{DiMatteo:2005p29098, Hopkins2005}, which show that  the supermassive black hole grows by accretion while enclosed by obscuring dust, and then sheds its obscuring cocoon, deposited by the merger, through outflows driven by powerful quasar winds \citep{Balsara:1993p28980}. This activity ceases when the fuel supply to the central regions is exhausted, starving both the active galactic nucleus (AGN) and the star forming activity. This scenario is also supported by the observational result that AGN activity increases with increasing bolometric luminosity \citep{Genzel:1998p29109, Tran:2001p29490, Lonsdale2006}.

In the low-redshift Universe, star formation is dominated by M82-type starburst galaxies, and less than 50 ULIRGs are known at $z<0.1$. However, at higher redshifts  the cosmic star formation rate is dominated by increasingly vigorous starburst galaxies. At $z=0.5$ ULIRGs are far more numerous than in the local Universe, and dominate the cosmic star formation rate \citep{Blain:2002p28984, LeFloch2005, Lonsdale2006, Wilman:2008}. They appear to represent a significant step in the evolution of galaxies, so that understanding their nature is crucial to understanding galaxy evolution.  However, the many magnitudes of extinction to their nuclei make it difficult to determine whether their dominant power source is accretion onto a supermassive central black hole, or a colossal starburst. The current consensus \citep[e.g.][]{Risaliti10, Veilleux09} is that most ULIRGs are primarily powered by a starburst, although an AGN is dominant in a significant fraction. However, this view is based largely on optical and near-infrared observations, which penetrate only the outer layer, and tell us little of the nucleus.

IRAS F00183-7111 (also known as IRAS 00182-7112, and which we shall hereafter refer to as 00183) is the most luminous ULIRG discovered in the IRAS survey, and one of the most luminous ULIRGs known. At a redshift of 0.3276 \citep{Roy:1997p29162}, it has a bolometric luminosity of $9 \times10^{12}$ \Lo \citep{Spoon:2009p29237}, most of which is radiated at far-infrared wavelengths, and a linear scale of 1 mas = 4.7 pc ($H_{0} = 71$ km s$^{-1}$ Mpc$^{-1}$). It is the most luminous of the extreme ULIRGs studied by \citet{Armus:1987p29090}, and also has the largest FIR excess,  with $L_{\rm{FIR}} / L_{\rm{B}} = 360$.  For comparison, Arp 220 has $L_{\rm{FIR}} / L_{\rm{B}} = 150$, and spectacular starburst systems like M82 and NGC 3256 have $L_{\rm{FIR}} / L_{\rm{B}} = 10$, so this is a heavily obscured object. Near-infrared imaging by \citet{Rigopoulou:1999p29130} (Figure \ref{fig:figIR}) appears to show a disturbed morphology and a single nucleus, although the many magnitudes of extinction, even at near-IR wavelengths, mean that this image may be showing an outer shell of the galaxy, rather than the nucleus. 
CO emission has also been detected from 00183 with an inferred molecular gas mass $> 2.4 \times 10^{10} $ \Msun \citep{Norris11}, confirming that this object is a vigorous star-forming galaxy. 

Optically \citep{Drake:2004p30337}, it appears as an unremarkable smudge 20 arcsec across \citep{Heckman1990}, but long-slit spectroscopy  shows bright ionised gas extending 50 kpc east and 10 kpc west of the nucleus, with highly disturbed kinematics.  Line widths are $> 600$ km s$^{-1}$ over a 30 kpc-diameter region, reaching 1000 km s$^{-1}$ in the region 3 - 7 kpc east of 
the nucleus.  Line profiles are double-peaked with a splitting of 300 - 400 km s$^{-1}$ in the region 10 kpc east, and the profiles are strongly blue-shifted with respect to the nucleus out to 25 kpc E.  The line ratios and widths are typical of Sy2 galaxies, and the kinematics suggest a powerful outflow. 

\cite{Spoon:2009p29237}  observed unusually wide [Ne II] 
and [Ne III] profiles in 00183 -- the widest [Ne II] 
and second widest [Ne III] lines in a sample of  
82 ULIRGs observed with Spitzer-IRS \citep{Spoon09}.
The lines may trace strongly disturbed gas, resulting from
interaction of the radio jets with the ISM. However, the strong dust extinction means that these are characteristics of the outer parts of the galaxy, rather than the nucleus. The size, velocity, and morphology of the outflow agrees well with the super-wind model of \citet{Heckman1990}, in which the optical emission comes from a thin shell of shock-heated ambient gas swept up by a wind driven by the overpressure produced by the starburst.

\citet{Spoon:2004p28328, Spoon2007} fail to detect several common mid-infrared AGN tracers in their Spitzer-IRS high resolution spectrum. Furthermore, \citet{Spoon:2004p28328} detect a deep 9.7 $\mu$m silicate absorption feature, and  strong  4.7 $\mu$m  CO absorption. All of these are unusual for classical Seyfert galaxies and quasars \citep{Spoon2007, Hao:2007p29931} suggesting that the line of sight to the nucleus is heavily obscured.

However, radio observations are unaffected by the heavy dust obscuration that blocks our view at shorter wavelengths, and allow us to see into the nucleus to look for typical AGN-related core-jet structures with high brightness temperature, to distinguish from diffuse, lower $T_{\rm{b}}$ starburst emission. At the centre of 00183 lies a strong radio source \citep{Roy:1997p29162}, whose radio luminosity is ten times higher than would be expected from a star-forming ULIRG of 10$^{13}$ \Lo. Instead, this radio luminosity is attributable to a radio-loud AGN buried at the core of 00183, with a radio luminosity of $L_{4.8GHz}=3\times10^{25}$ W Hz$^{-1}$ \citep{Roy:1997p29162} which places it within the regime of high luminosity (FRII-class) radio galaxies. This AGN is invisible at optical and near-infrared wavelengths because of the dense dust galaxy surrounding it, evidence for which includes  the deep 9.7 $\mu$m silicate absorption feature \citep{Tran:2001p29490, Spoon:2004p28328}. However, the AGN is confirmed by the detection of a 6.7 keV FeK line (Fe XXV) with a large equivalent width, indicative of reflected light from a Compton thick AGN \citep{Nandra:2007p29707}. 

In a sensitive 12-hr ATCA synthesis observation of 00183 at 4.8 GHz, Roy et al. \citep[unpublished data cited by ][]{Roy:1997p29162} found an unresolved core of 108 mJy and no lobes brighter than 30 K ($3 \sigma$).  The core has an extraordinary radio luminosity of $L = 3\times10^{25}$ W Hz$^{-1}$, typical of powerful radio galaxies, which makes the object $\approx 40$ times more radio-loud than expected for the normal radio-FIR correlation.  To produce this radio luminosity using supernovae, as in Arp 220, would require an implausible  20000 luminous radio supernovae (i.e. similar to RSN 1986J) radiating simultaneously.

However, the ultimate distinction between radio-loud AGN and star formation is best achieved using Very Long Baseline Interferometry (VLBI). While clumps of radio supernovae can emulate low-luminosity AGN, high luminosity VLBI detections are nearly always produced by AGN \citep{Kewley:2000p30027}. Here we 
present the results of a VLBI experiment to probe the radio core of 00183. 

\section{Observations and Data Reduction}

The VLBI observations took place (project code V024a) over a period of fifteen hours on 11 August 2000. Six radio telescopes were used for this experiment: the 64 m antenna of the Australia Telescope National Facility (ATNF) near Parkes; the Tidbinbilla 70-m DSS-43 Deep Space Network antenna at the Canberra Deep Space Centre, a phased array at the ATNF Australia Telescope Compact Array (ATCA) near Narrabri; the ATNF Mopra 22 m antenna near Coonabarabran; and the University of Tasmania's 26 m antenna near Hobart and their 32 m antenna at Ceduna.  The Narrabri phased array consisted of five 22 m antennas of the Australia Telescope Compact Array configured in a compact (EW750a) configuration, giving a collecting area equivalent to a 49 m diameter antenna. The Hartebeestehoek antenna in South Africa also took part in the observations, but no reliable data were obtained using that antenna, and so the Hartebeestehoek data have been omitted from the data reduction. Observing scans were centred on the nucleus of IRAS F00183-7111 ($\alpha = 00\fh20\fm34\fs65$; $\delta = -70\fd55\fm26\farcs4$ [J2000]) which doubled as a phase reference source. Hourly scans of IRAS F00183-7111 were made every two hours at 13 cm.

The observation used the S2 recording system \citep{Cannon:1997p10533} to record $2\times16$ MHz bands (digitally filtered 2 bit samples). Both bands were upper side band and right circular polarisation. The frequency ranges used for the two bands were 2252 - 2268 MHz and 2268 - 2284 MHz. The recorded data were correlated using the ATNF Long Baseline Array (LBA) processor at ATNF headquarters in Sydney \citep{Wilson:1992p9290} with the nominal system temperatures applied for each antenna. The data were correlated using an integration time of 5 seconds and with 32 frequency channels across each 16 MHz band (channel widths of 0.5 MHz). 

The correlated data were imported into the AIPS\footnote{The Astronomical Image Processing System (AIPS) was developed and is maintained by the National Radio Astronomy Observatory, which is operated by Associated Universities, Inc., under co-operative agreement with the National Science Foundation} package for initial processing. The data were fringe-fit (AIPS task FRING) using a one minute solution interval, finding independent solutions for each of the 16 MHz bands.

The fringe-fitting solutions were applied to the data and the resulting visibilities exported into DIFMAP \citep{Shepherd:1994p10583} for further calibration and imaging. The large-scale structure of the source was modelled in $u-v$ space with two Gaussian components and the smaller-scale structure with two delta functions. Natural weighting was used to maximise sensitivity and to accentuate the larger-scale structure of the source. Several iterations of model-fitting and phase self-calibration were executed until the modelled and observed visibilities converged. In the final iteration, amplitude self-calibration was performed with a solution period of 10 minutes. A one sigma rms noise of 90 \uJy beam$^{-1}$ was achieved, closely matching the theoretical thermal noise of the observation\footnote{Estimated with the ATNF VLBI sensitivity calculator: http://www.atnf.csiro.au/vlbi/calculator}. 
Errors of $\pm10$\% are listed due to uncertainties in the absolute flux density scale for Southern Hemisphere VLBI \citep{rey94}.

\begin{figure}
\epsscale{1.0}
\plotone{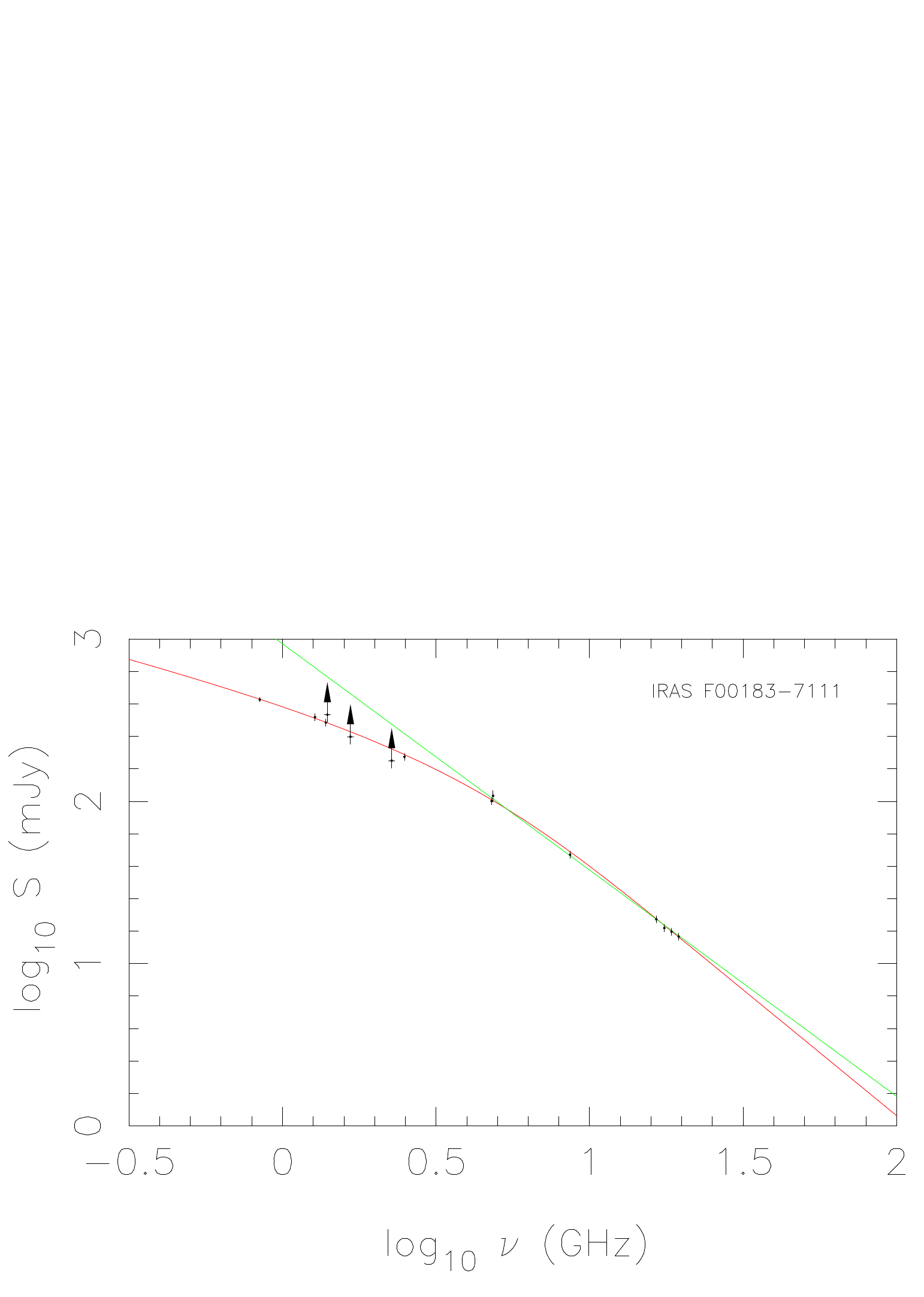}
\caption{Radio spectrum of 00182. Data used to construct this plot are listed in Table 1. The spectral index at high frequency is approximately constant  and flattens at low frequencies. The light (green) line shows the best-fit spectrum law with a  constant spectral index of $\alpha$ = -1.38 (where $I(\nu) \propto \nu^{\alpha}$). The dark (red) line shows a fitted GPS model \citep{Snellen:1998} with spectral index changing from -1.49 at high frequency to -0.43 at low frequency.}
\label{fig:figsed}            
\end{figure}

In addition, ATCA data on 00183 were taken from the ATCA archive and reprocessed to yield a flux density  at a number of frequencies. These, together with radio measurements from the literature, are listed in Table 1,  and are used to produce the radio spectrum shown in Fig. \ref{fig:figsed}. 

\section{Results}

\begin{figure}
\epsscale{1.0}
\plotone{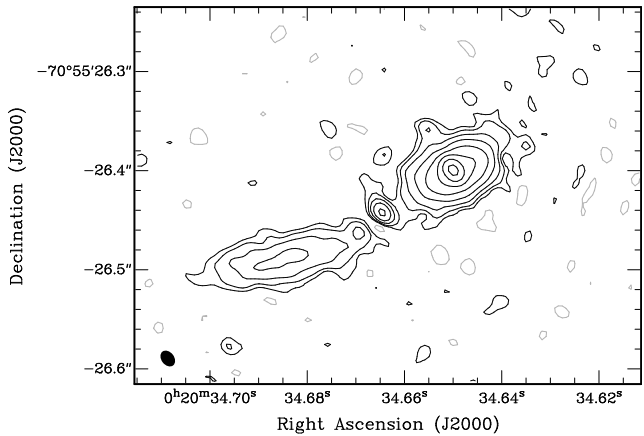}
\caption{Naturally-weighted 2.3 GHz LBA image of IRAS F00183-7111. The peak surface 
brightness is 45 mJy beam$^{-1}$ and the $1\sigma$ rms image noise is 90 \uJy beam$^{-1}$. 
Contours are drawn at $\pm1, \pm2, \pm4, \pm8, \cdots$ times the $3\sigma$ rms noise. The beam 
size is $12.8\times17.6$ mas at a position angle of $38\deg$.}
\label{fig:figlba}            
\end{figure}

The resulting VLBI image is shown in Figure \ref{fig:figlba}.

\begin{figure}
\epsscale{0.7}
\plotone{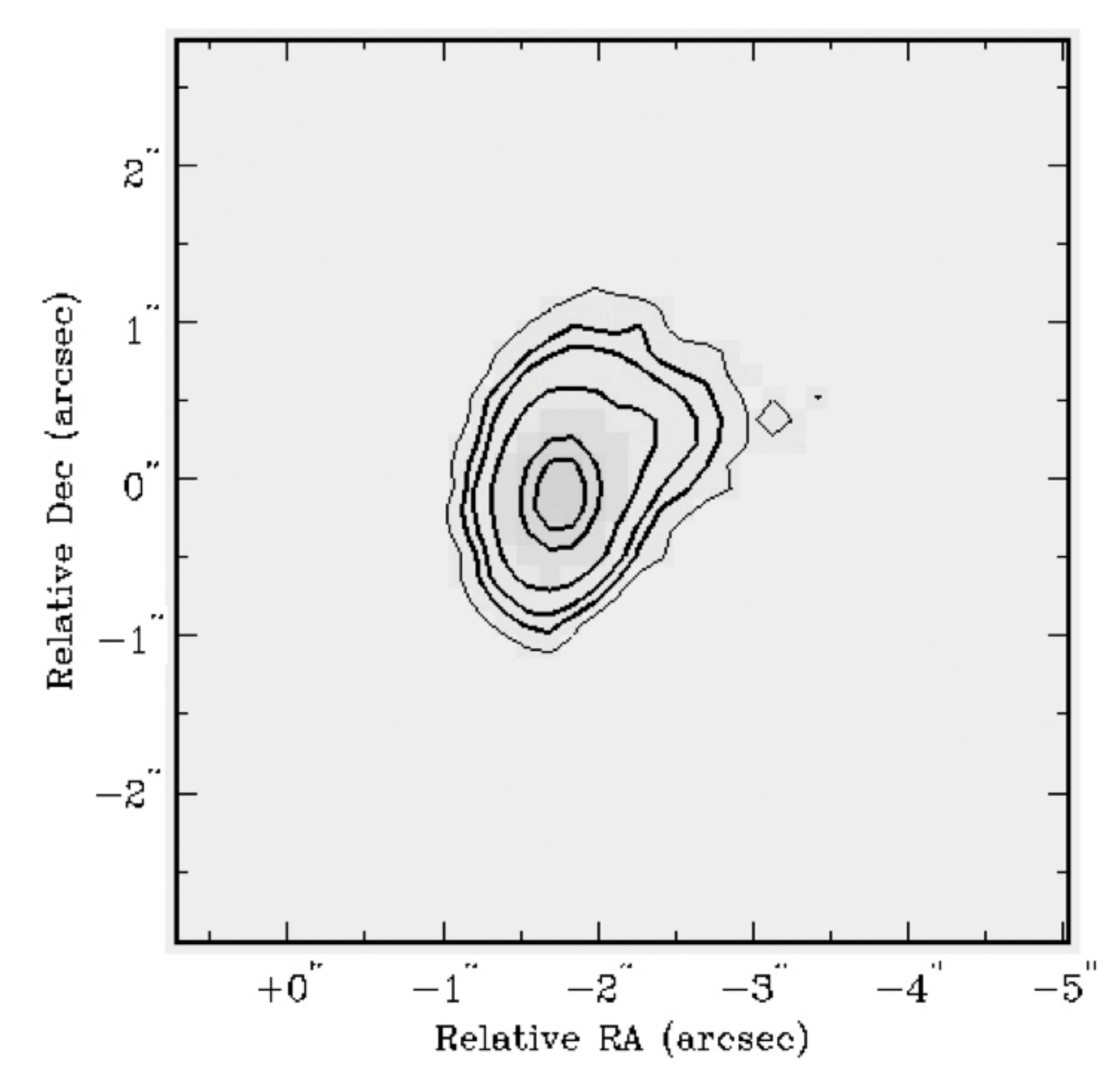}
\caption{Near-IR (2.16 $\mu$m) image of IRAS F00183-7111 adapted from \citet{Rigopoulou:1999p29130}. The orientation of this image was incorrect in the version published by Rigopoulou et al. and has been corrected here.}
\label{fig:figIR}            
\end{figure}

\begin{figure}
\epsscale{0.7}
\plotone{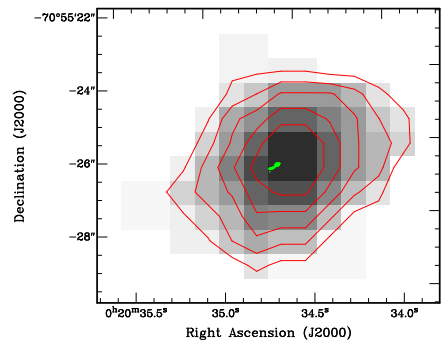}
\caption{UKST R-band image of IRAS F00183-7111 with contours (red) overlaid with naturally-weighted 2.3 GHz LBA contour map (green).}
\label{fig:figopt}            
\end{figure}
 
The VLBI image shows the classical double-lobed morphology similar to that of a classical radio galaxy, but on a  scale (1.7 kpc long) that is several orders of magnitude smaller, and therefore indicative of a Compact Steep Spectrum (CSS) source \citep{ODea:1998p10573, Randall2011}.

The jet is approximately in the same position angle as the extensions to the optical and infrared images shown in Figs. \ref{fig:figIR} and \ref{fig:figopt}, but higher-resolution images are needed to see whether this alignment is significant.  \citet{Drake:2004p30337} present a deeper R-band image showing an additional component to the east, and there is marginal evidence for soft X-ray emission to the east of the nucleus \citep{Nandra:2007p29707}.

The measured integrated flux density of the VLBI image (178 mJy) is close to the interpolated flux density (192 mJy) derived by assuming a constant spectral index between the 4.8 GHz measurement by \citet{Roy:1997p29162} and the 843 MHz measurement by \citet{Mauch:2003p28968}. Since \citet{Roy:1997p29162} also detected no extended emission, we surmise that the flux calibration scale is likely to be approximately correct, and that essentially all the flux from the source is contained within the image shown in Fig.\ref{fig:figlba}.

We note that the mid-point of the VLBI source does not fall on the peak, but slightly to the East of it, in a gap between the main peak and a subsidiary peak at the base of the Eastern jet. While it is tempting to speculate that these two peaks might represent the nuclei of the two galaxies that merged to form this ULIRG, it is also likely that the Eastern peak is simply a knot in the jet, and that the source is asymmetrical. Higher-resolution, higher-dynamic range observations will be necessary to confirm this.

\section{The Evolutionary State of 00183}
\subsection{AGN or Starburst?}

The literature contains conflicting evidence as to whether the radio emission from 00183 is primarily generated by star formation or an AGN. 

On one hand, 00183 is a classical ULIRG, and ULIRGs are generally assumed to be predominantly powered by starburst activity. Its infrared spectrum lacks the lines (7.65 $\mu$m [NeVI], and 14.32 \& 24.32 $\mu$m [NeV]) characteristic of an AGN \citep{Spoon:2009p29237}.  
It also  has an inferred molecular gas mass $> 2.4 \times 10^{10} $ \Msun \citep{Norris11}, and such high molecular masses are generally found only in vigorous star-forming galaxies. 
Furthermore, the strength of the  11.2 $\mu$m PAH feature   implies an energetic  starburst that may
contribute up to 30\% of the bolometric luminosity of the source \citep{Spoon:2004p28328}.

On the other hand, the radio luminosity of 00183 is $L = 3\times10^{25}$ W Hz$^{-1}$, typical of powerful radio galaxies, and placing it roughly on the break between Fanaroff-Riley Type I and Type II galaxies \citep{Martinez2006}. As discussed in \S\ref{intro}, this radio luminosity would require a physically unrealistic number of radio supernovae to produce it. Furthermore, the radio-infrared ratio of this object is $\approx 40$ times higher than expected for a star-forming galaxy, based on the canonical radio-FIR correlation 
 \citep{vanderKruit71,Helou93, deJong85,  Condon92, Mao2011}. 

For these and other reasons, \citet{Roy:1997p29162} and \citet{Spoon:2009p29237} argue that 00183 is powered by an AGN, which is obscured by a thick veil of dust, so that optical and near-IR observations see only the star-formation activity on the outside surface of the galaxy.

Figure \ref{fig:figlba} provides clinching evidence for this argument. The morphology of the radio source in 00183 is clearly that of a standard core+jet radio-loud AGN.

\subsection{Quasar mode or radio mode?}

00183 is an extremely luminous ULIRG with vigorous star formation, and contains a radio-loud AGN at its centre. 
Such ULIRGs appear to be advanced mergers of gas-rich spirals, caught after their first peri-passage \citep{Toomre:1972p29509, Sanders:1988p29218, Veilleux:2002p29465}. The leading merger scenario \citep{Sanders:1988p29218} predicts ULIRGs to evolve into dusty quasars before settling down as moderate mass ellipticals \citep{Dasyra:2006p29091, Dasyra:2006p29094}. The above scenario is largely in agreement with merger simulations \citep[e.g.][]{DiMatteo:2005p29098, Hopkins2005}, which show the supermassive black hole growing by accretion, becoming a proto-quasar. It then sheds its obscuring cocoon, deposited by the merger, through outflows driven by powerful quasar winds \citep{Balsara:1993p28980}. The central regions will eventually be cleared of fuel, starving both the AGN and any star forming regions in the bulge, and activity will cease. 

This scenario is broadly consistent  with the ``quasar-mode''/``radio-mode'' model
\citep{Best2006, Croton2006, Hardcastle2007}, in which the two black holes of the merging galaxies grow both through coalescence and through the accretion of cold disc gas from the host galaxies, accompanied by vigorous starburst activity in the cool gas. In this efficient quasar-mode (or cold-mode) accretion, the infall of cold gas results in an accretion disk within a hot broad-line region, producing a rapid growth of the black hole mass. Further out, the accretion disk extends into  a dusty torus, while above and below the torus is the narrow-line region. 

During the quasar mode accretion, the unified model \citep[e.g.][]{Antonucci93, Barthel89} explains the differences between broad-line (e.g. quasars) and narrow-line objects (e.g. narrow-line radio galaxies) in terms of orientation of the line of site relative to the torus. A radio-loud quasar is only seen as a quasar if it is seen face-on; otherwise the nucleus is obscured by the torus and it classified as a radio galaxy. 
However, unified models do not explain the difference between radio-loud and radio-quiet objects. The factors that cause a black hole to become radio-loud or radio-quiet are currently unclear, but may include black hole mass \citep{Best2006}, black hole spin \citep{Sikora07}, magnetic field \citep{Ye05}, merger history \citep{Wilson95} or a combination of these factors. 

Also missing from this description is the merger status of the two original black holes. The timescale for the black hole merger is extremely uncertain \citep[e.g.][]{Komossa03, Wilson95, Khan11, Preto11, Burke11} and so it is not clear at what stage of the above process the black holes merge, nor what the large-scale effect, if any, of that merger will be. We note a possible secondary peak discussed in \S 3, but defer further speculation until we have a higher resolution, higher dynamic-range, VLBI image.

Eventually enough energy is deposited into the cool inflowing gas, both from radio jets and star formation, to heat and disrupt the gas, resulting in the cessation of star-forming activity and a much less efficient accretion mode known as radio-mode or hot-mode accretion. The end product of this process is an elliptical galaxy, with little or no star formation, whose central black hole slowly accretes the hot interstellar medium.

00183 appears to be a rare example of an object caught at an early phase of the process of forming a quasar, in the brief period after the merging starburst, as the black hole ramps up to ``quasar mode''  activity. The radio source has already grown to a luminosity $L_{4.8GHz}=3\times10^{25}$ W Hz$^{-1}$ \citep{Roy:1997p29162}. 

A parallel, observational, description is given by \cite{Spoon2007} who propose an evolutionary scenario in which 00183
(located in their class "2A") is in a short-lived evolutionary 
 transition from fully obscured buried nuclei to  an AGN with a clumpy torus
geometry. 

\subsection{PRONGS}

\citet{Norris:2006p6530} identified a class of galaxies, subsequently christened PRONGS (Powerful Radio Objects Nested in Galaxies with Star-formation) by \cite{Mao2010}, that consist of a radio-loud AGN buried within a galaxy that appears at optical/infrared wavelengths as a star-forming galaxy. Although AGNs are widely found within star-forming galaxies (e.g. Seyfert galaxies), such AGNs are relatively weak ($\sim 10^{21}$ W Hz$^{-1}$), three orders of magnitude weaker than PRONGS. Furthermore, Seyfert galaxies typically fall on the far-infrared-radio correlation \citep[FRC:][]{Roy98} whereas 00183 and other PRONGS are typically $\sim$ 10 times too radio-luminous to follow the FRC.

PRONGS are rare in the local Universe, with only two known examples \citep[NGC 612 \& 0313-192: ][]{Ekers:1978p30180, Emonts2008, Ledlow:2001p30193,Keel:2006p30187} but seem to be increasingly common at high redshifts, with $\sim 50$ identified by Mao et al. in the ATLAS survey of seven square degrees \citep{Norris:2006p6530, Middelberg2008}. The PRONGS studied so far are smaller ($< $50 kpc to 200 kpc) than standard radio galaxies, 
which range from hundreds of kpc to 4.5 Mpc\citep{Schilizzi01}, suggesting that they may be either young or frustrated radio-loud galaxies. 

00183 is clearly an extreme example of the class of PRONGS. If, as \citet{Mao2010} suggest, PRONGS  are an early evolutionary stage of radio-loud galaxies, then 00183 may represent a short-lived extreme phase at the start of this process. 

\subsection{GPS and CSS galaxies}
Gigahertz-Peaked Spectrum (GPS) and Compact Steep Spectrum (CSS) sources are similar to PRONGS in that they have a very small but powerful radio AGN at their centre. GPS sources are typically $<$  1 kpc in size, and are widely thought to mark the earliest evolutionary phase of large-scale radio sources \citep{Polatidis03, Tinti06, Fanti09}.  CSS sources are one to several tens of kpc in size \citep{ODea:1998p10573, Snellen:1998, Randall2011} and represent an intermediate stage between GPS sources and the largest radio sources, FRI/II galaxies \citep{Snellen99}. PRONGS differ from CSS/GPS sources  in that they are immersed in a star-forming galaxy, whereas most CSS/GPS sources are hosted by luminous red galaxies or quasars \citep{ODea:1998p10573, Randall2011}.  

CSS sources have a steep ($\alpha < -0.8$) spectral index across the GHz frequency range, while GPS sources reach a maximum brightness at a few GHz, with the spectral index becoming positive below the maximum \citep{Randall2011}. The positive spectral index at low frequencies is thought to be caused by synchrotron self absorption, although free-free absorption may also be significant \citep{Fanti09}. 
 Fig. \ref{fig:figsed} shows the spectral energy distribution (SED) of 00183. 00183 is intermediate between these two classes, with the spectrum noticeably flattening at low frequencies without actually reaching a maximum.
 
We can estimate whether free-free absorption could be causing the observed turnover in this source. We assume a constant synchrotron spectral index $\alpha = -1.38 $ (i.e. the value at high frequencies), and a flux density of 14.7 mJy at 19.513 GHz, which implies an un-absorbed
synchrotron flux density of 1120 mJy at 843 MHz.  If the turnover is caused by free-free absorption by the surrounding starburst, modelled as a foreground slab of ionised gas, then the  slab has an
opacity of $\sim$ 1 at 843 MHz. Assuming an electron temperature of 8000K, and an electron density of 
$10^{3}$ to $10^{4}$ cm$^{-3}$ \citep{Anant97}, the required path
length to produce the observed
spectral flattening is 1.5 to 15 pc, with a  covering factor of unity. This is reasonable given the intense starburst, and demonstrates that, at least in this source, the turnover can easily be caused by free-free absorption in the surrounding starburst.

In summary, the radio luminosity, size (1.7 kpc), and spectral shape mark 00183 as similar to CSS and GPS sources, but differing in that it is hosted by a star-forming ULIRG. According to the standard CSS/GPS scenario \citep{Randall2011}, the radio jets have just turned on, and are now boring their way through the dense dust and gas.  00183 is surrounded by an enormous mass of cold molecular gas (\Mgas $> 2.4 \times 10^{10} $ \Msun) \citep{Norris11} so clearly the jets have a difficult task ahead of them.
They will eventually emerge as fully-fledged giant radio lobes, at which point 00183 will probably become a classical FRII galaxy or quasar. 

\section{Conclusion}

Our VLBI image of the extreme ULIRG F00183-7111 shows  a classical core-jet radio-loud ($ > 10^{25}$ W Hz$^{-1}$ ) AGN, which has the luminosity and morphology of an FRII galaxy, except that the total length of the source is only 1.7 kpc. This core is surrounded by a vigorous starburst accompanied by some $ 2.4 \times 10^{10} $ \Msun\ of molecular gas. The many magnitudes of dust extinction  accompanying this starburst totally obscure any sign of this AGN at optical or infrared wavelengths.

We have placed this source in an evolutionary sequence defined by the standard quasar-mode/radio-mode model, and find that it represents the initial stage of the formation of a quasar. It is in the earliest throes of quasar-mode accretion, with the powerful radio jets still confined by the dense cool gas from the galaxies that merged to form this system. 

In this respect it is very similar to the classical CSS/GPS sources, differing only in that its host galaxy appears to be at an earlier stage of post-merger evolution than most CSS/GPS hosts.
Eventually the radio jets will drill their way out of the gas, at the same time heating and disrupting the gas. As a result,
 the current vigorous burst of star formation will be extinguished, marking the transition of this source from a star-forming ULIRG with a ``quasar-mode''  AGN at its centre to a powerful radio-loud quasar.

\section*{Acknowledgments}

We thank the Australian LBA team whose continuing efforts make observations like this possible, and especially the University of Tasmania for making the Hobart and Ceduna antennas available, and to the CDSCC for making the Tidbinbilla antenna available. We also wish to thank Associate Professor Hiroshi Imai (Kagoshima University) for kindly giving us access to 1 cm ATCA data for IRAS F00183-7111, and Sarah Burke-Spolaor for helpful comments on black hole mergers.

The Australia Telescope is funded by the Commonwealth of Australia for operation as a National Facility managed by CSIRO. This research has made use of the NASA/IPAC Extragalactic Database (NED) which is operated by the Jet Propulsion Laboratory, California Institute of Technology, under contract with the National Aeronautics and Space Administration.

\begin{table*}
\centering
\begin{minipage}{140mm}
\caption{Radio data used for Fig. \ref{fig:figsed}}
\begin{tabular}{lllllll}
\hline
Date                      & Obs.  & $\nu$  & Resolution & $S_{\nu}$ & Ref. \\
                          &       & (GHz)  & (arcsec)   & (mJy)     &      \\
\hline
1997 - 2003               & SUMMS & 0.843  & 45    & 423.7 & \citet{Mauch:2003p28968}	\\
08 Feb 2004               & ATCA  & 1.276  & 6.7   & 330   & This paper	\\
25 Jan 2002 - 31 Jan 2002 & ATCA  & 1.384  & 7     & 306   & \citet{Drake:2004p30349}	\\
20 Apr 2001 - 26 Apr 2002 & LBA   & 1.4    & $\sim0.1$  & 341.8 & \citet{Drake:2004p30349}	\\
                          & PTI   & 1.665  & 0.1   & 250   & \citet{Norris:1988p28994}   \\
11 Aug 2000               & LBA   & 2.268  & 0.015 & 178   & This paper   \\
25 Jan 2002 - 31 Jan 2002 & ATCA  & 2.496  & 4     & 188   & \citet{Drake:2004p30349}	\\
05 Jan 1998 - 28 Feb 1998 & ATCA  & 4.794  & 2     & 100   & \citet{Drake:2004p30349}	\\
                          & PMN   & 4.85   & 252   & 108   & \citet{Wright:1994p17797}	\\
05 Jan 1998 - 28 Feb 1998 & ATCA  & 8.64   & 1     & 46.9  & \citet{Drake:2004p30349}	\\
13 Jun 2009               & ATCA  & 16.489 & $1.0\times0.47$  & 18.7  & This paper	\\
13 Jun 2009               & ATCA  & 17.513 & $0.94\times0.46$ & 16.6  & This paper	\\
13 Jun 2009               & ATCA  & 18.489 & $0.89\times0.43$ & 15.7  & This paper	\\
13 Jun 2009               & ATCA  & 19.513 & $0.85\times0.41$ & 14.7  & This paper	\\
11 Jul 2009                & ATCA  & 86.84 & 28 & 3.4   & \cite{Norris11}	\\
\hline
\end{tabular}
\end{minipage}
\end{table*}

\label{lastpage}

\end{document}